\global\def\draftcontrol{0}
\makeatletter
\newbox\draftbox

\font\smallcomptr=cmtt8
\font\smallrm=cmr8
\def\strip#1>{}
\def\verbatim#1{\def\next{#1}%
  {\smallcomptr\expandafter\strip\meaning\next}}
\def\draftcite#1{\ifnum\draftcontrol=1\verbatim{#1}\else{}\fi}

{\count255=\time\divide\count255 by 60
\xdef\hourmin{\number\count255}
\multiply\count255 by-60\advance\count255 by\time
\xdef\hourmin{\hourmin:\ifnum\count255<10 0\fi\the\count255}}
\def\draftdate{\number\month/\number\day/\number\year\ \ \ \hourmin }
\def\ps@draft{\let\@mkboth\@gobbletwo
    \def\@oddhead{}
    \def\@oddfoot
       {\hbox to 7 cm{{\smallcomptr \jobname.tex}{\smallrm\quad \draftdate}
       \hfil}\hskip -7cm\hfil\rm\thepage \hfil}
    \def\@evenhead{}\let\@evenfoot\@oddfoot}

\let\l@bel=\label
\renewcommand{\label}[1]{\l@bel{#1}
  \global\setbox\draftbox=\hbox{$\quad$\draftcite{#1}}
  \wd\draftbox=0mm}
\renewcommand{\@eqnnum}{\rm{(\theequation)}\box\draftbox}

\def\@bibitem#1{\item\hskip -3cm \hbox to 2cm
{\hfil\draftcite{#1}}\hskip 1cm
\if@filesw \immediate\write\@auxout
       {\string\bibcite{#1}{\the\value{\@listctr}}}\fi\ignorespaces}

\makeatother

\newcommand{\da}{\partial}

\newcommand{\un}{\underline}

\newcommand{\hf}{{_1\over^2}}
\newcommand{\ov}{\overline}

\newcommand{\gl}{\theta}

\newcommand{\NR}{{{\bf R}}}
\newcommand{\NC}{{{\bf C}}}

\newcommand{\ii}{{\rm i}}

\newcommand{\CF}{{\cal F}}

\newcommand{\qq}{\begin{equation}}
\newcommand{\qqq}{\end{equation}}
\newcommand{\myeqnarray}[1]{
  \begingroup
  \jot=#1pt
  \arraycolsep=2pt
  \begin{eqnarray}}
\newcommand{\beqnarray}{\myeqnarray{3}}
\newcommand{\eeqnarray}{\end{eqnarray}\endgroup}
\newcommand{\ba}{\begin{array}{cc}}
\newcommand{\ea}{\end{array}}

\documentclass[12pt]{article}
\usepackage[xdvi]{graphicx}%
\usepackage{dcolumn}
\usepackage{epsfig}
\newtheorem{pro}{Proposition} 
\newtheorem{definition}{Definition}

\title{Large deviations of lattice  Hamiltonian dynamics coupled to stochastic thermostats.}
\date{\today}
\author{Thierry Bodineau\footnote{Ecole Normale Sup\'erieure, DMA, 45 rue d'Ulm
75230 PARIS Cedex 05, FRANCE}
 and Rapha\"el Lefevere\footnote{LPMA, Universit\'e Denis Diderot (P7) - Bo{\^ \i}te courrier 7012, 75251 PARIS Cedex 05, FRANCE}
}

\begin{document}

\maketitle

\begin{abstract}
We discuss the Donsker-Varadhan theory of large deviations in the framework of Hamiltonian systems thermostated by a Gaussian stochastic coupling.  
We derive a general formula for the Donsker-Varadhan large deviation functional for dynamics which satisfy natural properties under time reversal. 
Next, we discuss the characterization of the stationary states as the solution of a variational principle and its relation to the minimum entropy production principle.
Finally, we compute the large deviation functional of the current in the case of a harmonic chain thermostated by a Gaussian stochastic coupling.
\end{abstract}

\section{Introduction.}

In the recent years, several studies  
of large systems out of equilibrium through fluctuation theory have been made \cite{jona1,jona2, BodineauDerrida,BD2,EvansCohenMorriss,GallavottiCohen1,GallavottiCohen2, Kubo}.  
In a recent series of papers 
\cite{MaesNetocny1,MaesNetocny2,MaesNetocny3,MaesNetocny4,MaesNetocny5}, it has been understood that in random systems driven out of equilibrium, the theory of large deviations provides naturally  a variational characterization of the steady states which is related to the minimum entropy production principle.  
In this paper, we pursue this approach in the framework of thermostated lattices of Hamiltonian oscillators
and investigate in this setting the Donsker-Varadhan large deviation theory \cite{DonskerVaradhan, DonskerVaradhan2}.
The existence of a large deviation principle for a chain of Hamiltonian oscillators has been already established in 
\cite{Wu, rbt2}, here we focus on the physical interpretation of the functional in terms of entropy production and on exact computations for Gaussian dynamics.

\medskip

Thermostated lattices of Hamiltonian oscillators arise naturally in physics as they model either the contact with an environment, the nonlinearity of the dynamics or the randomness of initial conditions by effective  stochastic Gaussian terms added to Hamilton's equations of motion.  For the class of models considered in this paper and introduced in section \ref{sec: examples}, the noise is degenerate as it acts only on some coordinates (the momenta).
Building on \cite{Lefevere}, we introduce dynamics with a deterministic driving force which are analogous to the asymmetric dynamics in lattice gas models.
These models satisfy the {\it generalized detailed balance} which is a relation between the generator of the dynamics, a reference measure and a time-reversal breaking function that is a linear combination of the local energy currents. 

\medskip

In section \ref{sec: DV}, we rephrase the Donsker-Varadhan theory in our setting \cite{Wu} and establish a useful formula (\ref{lemma1}) for the large deviation functional of processes which have natural properties under time reversal.
This generalizes  the formula expressing the functional as the Dirichlet form of the process in the case of reversible dynamics.

In section \ref{subsec: Entropy production}, we discuss, along the lines of 
\cite{MaesNetocny1,MaesNetocny3,MaesNetocny4,MaesNetocny5}, the  characterization of the stationary measures out of equilibrium in terms of a variational principle  and we stress the differences and the peculiarities inherent to the Hamiltonian structure.  
In particular, Hamiltonian dynamics leaves the Gibbs (or Shannon) entropy invariant, thus it comes as no surprise that the so-called minimum (Gibbs) entropy production principle fails to determine correctly stationary states in systems out of equilibrium  or even in equilibrium: it is blind to the Hamiltonian dynamics.  The only part of the dynamics modifying entropy is the diffusion modeling the action of the thermostats which, in our case, act only on the momenta variables.  
The degeneracy of the dynamics (non-ellipticity) is the main reason for which  the minimum entropy production fails to select the correct stationary state. We remark that in \cite{MaesNetocny2} (section 3.3),
this principle was shown to be valid in the case of an Hamiltonian system coupled to a diffusion acting on 
{\it all} the variables.
We analyze the interplay between the entropy producing part of the dynamics and the purely Hamiltonian part and show that the Donsker-Varadhan functional provides a variational characterization of the stationary measure (both in equilibrium and out of equilibrium) related to the minimum entropy production principle.

Finally in section \ref{subsec: fluctuation theorem}, we recover from  the symmetries of the  Donsker-Varadhan functional
the Gallavotti-Cohen symmetry \cite{GallavottiCohen1,GallavottiCohen2,EvansCohenMorriss,Kurchan, Kurchan2,LebowitzSpohn, Maes,Eckent,rbt2}.

\medskip

In section \ref{sec: Large deviations of the current}, we focus on  the large deviations of the heat current for 
thermostated lattices of harmonic oscillators. We compute exactly the functional and relate it to previous
expressions derived for lattice gas models \cite{jona2, BodineauDerrida,BD2}.

\section{Models}  
\label{sec: examples}

We will first recall the general framework of Hamiltonian dynamics and define the relevant physical quantities in this context. 
In section \ref{subsec: The generalized detailed balance}, the notion of {\it generalized} detailed balance is introduced.
It will be an important feature of the Hamiltonian systems with a stochastic forcing considered in this paper (see  sections \ref{eq: Heat in the bulk}, \ref{eq: Heat at the boundary} and \ref{sec: asymmetric}).

\subsection{Hamiltonian dynamics} 
\label{eq: Hamiltonian dynamics}

We consider a one-dimensional lattice of $N$  particles $(\un q,\un p)=(q_i,p_i)_{1\leq i\leq N}$, each one moving around an equilibrium position, the position of the $i$-th particle is denoted by $q_i$ and its momentum by $p_i$.  The systems we have in mind are described by a Hamiltonian, which is the energy function of the set of particles,
\qq
H(\un q,\un p)=\sum_{i=1}^N \left[\frac{p_i^2}{2}+V(q_i)+ \hf \big( U(q_{i+1}-q_i) + U(q_i-q_{i-1}) \big) \right].
\label{Hamilton}
\qqq
We will consider either periodic boundary conditions (with the convention $q_{N+1} = q_1$ and $q_0 = q_N$) or open systems with 
 the convention $q_0 = q_{N+1} = 0$.
 $V$ is the potential energy corresponding to an interaction with an external substrate.  $U$ describes the potential energy of the interaction between nearest-neighbours.  
Precise assumptions on the growth of the potentials will be detailed in Section \ref{subsec: DV functional} along with the mathematical statements.

\bigskip

The  positions $q_i$ and momenta $p_i$ of the particles obey the Hamilton's equations,
\begin{eqnarray}
\label{eq: Hamilton dynamics}
\dot q_i=\frac{\partial H}{\partial p_i},
\qquad \qquad
\dot p_i=-\frac{\partial H}{\partial q_i} \, .
\end{eqnarray}
The generator of the Hamiltonian dynamics is given by
\qq
L_H=\sum_{i=1}^N -\frac{\partial H}{\partial q_i}\frac{\partial}{\partial p_i}+\frac{\partial H}{\partial p_i}\frac{\partial}{\partial q_i} \, .
\label{eq: LH}
\qqq

\bigskip

In order to describe the propagation of heat in the lattice, one defines a local energy function,
\qq
h_i(\un p,\un q)=\frac{p_i^2}{2}+V(q_i)+\hf(U(q_{i}-q_{i+1})+U(q_{i-1}-q_{i})),
\label{local}
\qqq
such that  $H=\sum_{i=1}^N h_i$.  
The local energy transfer is identified with the transfer of mechanical energy between nearest-neighbours.
The energy current is therefore defined through the time evolution of the local energy,
\qq
\frac{d h_i}{dt}= L_H h_i = j_i - j_{i-1}
\label{conservation0}
\qqq
with $j_i$ the microscopic current of energy or heat between oscillator $i$ and $i+1$
\qq
j_i= - \hf U' (q_{i}-q_{i+1})(p_i+p_{i+1}).
\label{current0}
\qqq
In section \ref{sec: Large deviations of the current}, we will investigate the large deviations of the spatial average of the  current defined by
\begin{eqnarray}
\label{eq: current}
J (\un q,\un p) =  {1 \over N} \sum_{i = 1}^N j_i \, .
\end{eqnarray}

When such a Hamiltonian system is in thermal equilibrium at a temperature $T = \beta^{-1}$, its statistical properties are described by the Boltzmann-Gibbs  probability distribution over the phase space $\NR^{2N}$,
\begin{eqnarray}
\label{eq: Gibbs}
\rho (\un p, \un q)= {1 \over Z} \exp \left( -  {1 \over T} H (\un p, \un q) \right).
\end{eqnarray}
Remark that with respect to that distribution, 
$\left <p_i^2\right>= T$ and the averaged current is null $\left<j_i \right>=0$.  This last identity is straightforward because the equilibrium distribution is even under the reversal of momenta $\un p\rightarrow -\un p$ while the current is odd.

\subsection{The generalized detailed balance}
\label{subsec: The generalized detailed balance}

We now recall the notion of reversibility in the framework of  lattices of coupled oscillators. Then, we introduce 
the notion of {\it generalized detailed balance} which will apply when the chain of oscillators is forced out of equilibrium
by stochastic heat baths (sections \ref{eq: Heat in the bulk}  and \ref{eq: Heat at the boundary}) or a driving force (section \ref{sec: asymmetric}).

\bigskip

In Hamiltonian dynamics, the variables $\un p$ basically describe the velocities of the particles and are therefore odd functions under time reversal.
When isolated, those systems are reversible in the sense that if one lets evolve the particles from some initial conditions at time $0$ up to some time $T$ and then reverse all momenta,  Hamilton's dynamics will take back the particles through the same trajectory but with reverse momenta.   
Thus in the case of Hamiltonian dynamics (with possibly a stochastic forcing), the standard notion of detailed balance has to take into account the reversal of momenta. 
We denote by $\Pi$ the operator which reverses momenta
\begin{eqnarray}
\Pi (\un q,\un p)  =  (\un q, - \un p) \, .
\label{eq: Pi}
\end{eqnarray}
Let $P_t \left( (\un q,\un p), (\un q',\un p') \right)$ be the semi-group associated to an Hamiltonian dynamics with a stochastic perturbation (for a precise definition see sections \ref{eq: Heat in the bulk}  and \ref{eq: Heat at the boundary}) with initial data $( \un q,\un p) $ and  final data $(\un q',\un p')$ at time $t$.
We consider a probability  measure with density $\rho(\un q,\un p)$ symmetric  wrt to time reversal $\Pi \rho=\rho$.
The dynamics is  reversible wrt to the density $\rho(\un q,\un p)$ if for any time $t$ 
\begin{eqnarray}
\rho  \left( \un q,\un p \right)  \;  P_t  \left( (\un q,\un p), (\un q',\un p') \right)  = 
\rho  \left(  \un q',\un p'  \right)  \;  P_t  \left( \Pi (\un q',\un p'), \Pi (\un q,\un p) \right)   \, .
\label{eq: detail 0}
\end{eqnarray}
Let $L$ be the generator of this dynamics and let $L^\dagger$ denote the adjoint with respect to the Lebesgue measure. The adjoint $L^*_\rho$ of the operator $L$ with respect to the 
measure with density $\rho(\un q,\un p)$ is defined by  
\qq
L^*_\rho f(\un q,\un p) =\rho^{-1} (\un q,\un p) \; L^\dagger \big( \rho (\un q,\un p) f (\un q,\un p) \big)
\label{eq: adjoint}
\qqq
for any regular function $f(\un q,\un p)$.
Alternatively, $L^*_\rho$  satisfies
\begin{eqnarray*}
\left<fLg\right>_\rho=\left<g L^*_\rho f\right>_\rho \, .
\end{eqnarray*}
$L^*_\rho$ is therefore interpreted as the generator of the time reversed dynamics sampled with initial data distributed according to $\rho$. 
Applying (\ref{eq: detail 0}) for an infinitesimal amount of time, the equivalent form of the detailed balance
relation can be obtained
\begin{eqnarray}
L^*_\rho=\Pi L\Pi\, ,
\label{eq: detail 1}
\end{eqnarray}
where we used that the density $\rho$  satisfies $\Pi\rho=\rho$. 
For stochastic dynamics, the usual detailed balance relation does not involve the time reversal operator, however we keep the same terminology for simplicity.

\vskip.5cm

For general Hamiltonian systems coupled to stochastic thermostats, the reversibility may not hold and
(\ref{eq: detail 1}) has to be replaced by the {\it generalized detailed balance relation} (see for example \cite{Eckent}) which can be defined as follows 
\begin{definition}
\label{def: generalized detailed balance}
We consider an evolution with generator $L$ and $\rho(\un q,\un p)$  a reference measure such that $\Pi\rho=\rho$.
If there exists a function $\sigma (\un q,\un p)$  such that 
\begin{eqnarray}
\label{eq: sigma operateur}
L^*_\rho=\Pi L\Pi+\sigma, \qquad {\rm and} \qquad  \Pi \sigma =-\sigma \,,
\end{eqnarray}
then the triplet $(L,\rho,\sigma)$ is said to satisfy a generalized detailed balance relation. 
By convention in (\ref{eq: sigma operateur}), $\sigma$ acts as a multiplication operator.
\end{definition}

This definition will be illustrated in the examples introduced in sections \ref{eq: Heat in the bulk}, \ref{eq: Heat at the boundary} 
and  \ref{sec: asymmetric}.
In equilibrium dynamics (when $\sigma=0$) and when sampled with initial conditions distributed according to the equilibrium measure, the dynamics is equivalent to its time-reverse, this is the detailed balance principle.  When the system is subject to some non-equilibrium dynamics forcing a heat current through the system,  the stationary state looses its invariance under time-reversal ($\sigma \not = 0$).  In Markovian systems, 
the generalized detailed balance is a central feature of the Gallavotti-Cohen  symmetry \cite{Maes, EvansCohenMorriss, GallavottiCohen1,Kurchan,LebowitzSpohn,rbt, rbt2} for the large deviation functional of the function $\sigma$ (see section \ref{subsec: fluctuation theorem}).

Typically $\sigma$ is a function of the microscopic currents of energy in the lattice and the reference measure is an equilibrium or a local equilibrium distribution.  A given dynamics may satisfy a generalized detailed balance relation with respect to different reference measures and  different functions $\sigma$ (see section \ref{eq: Heat at the boundary}).

\subsection{Heat in the bulk}
\label{eq: Heat in the bulk}

The simplest perturbation of the Hamiltonian dynamics (\ref{eq: Hamilton dynamics}) is to couple each oscillator to a heat bath.  
The  boundary conditions are fixed $q_0 = q_{N+1} = 0$ and  each oscillator $i=1,\ldots,N$ evolves according to
\beqnarray
\label{eq: dynamics bulk}
dq_i&=&p_i dt  \\
dp_i&=&- \gamma_i p_i dt-\frac{\partial H}{\partial q_i} dt+\sqrt{2\gamma_i T_i} dw_i \, ,  \nonumber
\eeqnarray
where the $w_i$ are standard independent Brownian motions, $T_i = \beta_i^{-1}$ is the temperature of each heat bath and
$ \gamma_i  > 0$ is the friction.
The generator of the dynamics is given by
\qq
L=L_S+L_H
\qqq
where the Hamiltonian part $L_H$ was introduced in  (\ref{eq: LH}) and the symmetric part is
\qq
L_S= \sum_{i=1}^N  - \gamma_i p_i \frac{\partial}{\partial p_i} + \gamma_i T_i \frac{\partial^2}{\partial p_i^2} \, .
\label{eq: LS}
\qqq
When the temperatures $\{ T_i \}_i$ are not equal, the reversibility is lost and the invariant measure unknown.
The self-consistent chain \cite{BLL}, for which the temperatures $\{ T_i \}_i$ are tuned in order to maintain a constant 
average microscopic current along the chain, falls in the framework of the dynamics (\ref{eq: dynamics bulk}).

\bigskip

We check now that this dynamics satisfies the generalized detailed balance (see definition
\ref{def: generalized detailed balance}).
Given a collection of inverse temperatures $\un\beta=(\beta_1,\ldots,\beta_N)$, we 
take as a reference measure the following density 
\qq
\rho_{\un\beta} (\un p, \un q) ={ 1 \over Z} \exp \left( - \hat H (\un p, \un q) \right),
\quad {\rm with}  \quad
\hat H (\un p, \un q) = \sum_{i=1}^N \beta_i \, h_i (\un p, \un q) \, ,
\label{ref0}
\qqq
where $h_i$ are the local energies  introduced in (\ref{local}) and $\beta_i^{-1}=T_i$ is the temperature of the heat bath at site $i$.
We compute now $L^*_{\rho_{\un\beta} }$ the adjoint of $L$ with respect to the measure $\rho_{\un\beta}$.
Since $L_S = \Pi L_S \Pi $ and the stochastic part of the dynamics is reversible with respect to the measure $\rho_{\un\beta}$, one has for any function $f$ 
\qq
L_S f (\un q,\un p) 
= \rho^{-1}_{\un\beta} (\un q,\un p)  L_S^\dagger  \big( \rho_{\un\beta} (\un q,\un p) f (\un q,\un p) \big) \, .
\qqq
We turn now to the Hamiltonian part. 
Thanks to the relation $L_H^\dagger = - L_H$, we get 
\qq
\rho^{-1}_{\un\beta}  L_H^\dagger \rho_{\un\beta} =\sum_{i=1}^N \beta_i L_H h_i \, .
\qqq
From  (\ref{conservation0}), one has $L_H h_i=j_i - j_{i-1}$ with the convention $j_0=j_N=0$, thus 
\beqnarray
\label{eq: LH hat H}
\rho^{-1}_{\un\beta}  L_H^\dagger \rho_{\un\beta}  = \sum_{i=1}^N \beta_i (j_i - j_{i-1}) 
= - \sum_{i=1}^{N-1} (\beta_{i+1} - \beta_i) j_i \, .
\eeqnarray
Since $\Pi L_H\Pi=-L_H=L_H^\dagger$,  this implies that
\begin{eqnarray}
L^*_{\rho_{\un\beta}} = \Pi L\Pi+\sigma_{\un\beta}, \qquad {\rm with}  \quad
\sigma_{\un\beta}= - \sum_{i=1}^{N-1} (\beta_{i+1}-\beta_i) j_i \, .
\label{eq: sigma 1}
\end{eqnarray}
Thus the triplet $(L,\rho_{\un\beta},\sigma_{\un\beta})$ satisfies a generalized detailed balance relation
and $\sigma_{\un\beta}$ is a linear combination of the local currents.

\subsection{Heat at the boundary} 
\label{eq: Heat at the boundary}

If there are no heat baths in the bulk ($\gamma_i = 0$ for $ i=2, \ldots,N-1$), the dynamics (\ref{eq: dynamics bulk}) represents a crystal of atoms heated at two different temperatures at its boundaries. 
The equations of motion are given by,
\beqnarray
dq_i&=&p_idt,\quad i=1,\dots,N,
\label{eqq}
\\
dp_i&=&-\frac{\da H}{\da q_i}(\un p,\un q)dt,\quad i=2, \ldots,N-1,
\nonumber
\eeqnarray
and,
\begin{eqnarray}
dp_{1}&=&-\frac{\da H}{\da q_{1}}(\un p,\un q)dt-\gamma p_{1} dt
+\sqrt{2\gamma T_1}\,dw_1 \, ,  \label{eqp2} \\
dp_{N}&=&-\frac{\da H}{\da q_N}(\un p,\un q)dt-\gamma p_{N} dt
+\sqrt{2\gamma T_N}\,dw_N \, , \nonumber
\end{eqnarray}
where $T_1$  and $T_N$ stand for the temperature of the left and
right reservoirs, respectively,
whereas  $w_1$ and $w_N$ are two independent standard Wiener processes.

When $T_1=T_N=T=\beta^{-1}$,
the Gibbs measure (\ref{eq: Gibbs}) is invariant (stationary) for the stochastic dynamics defined above.
For two different temperatures, existence, uniqueness and
exponential convergence to an unique invariant state has been established under fairly
general conditions on the potentials $U$ and $V$ \cite{Carmona,eck1,eck3, rbt}.

\bigskip

A computation similar to the case of heat baths in the bulk (\ref{eq: sigma 1}) shows that
the dynamics (\ref{eqq}) satisfies the generalized detailed balance (see definition
\ref{def: generalized detailed balance}) for  any reference measure 
$\rho_{\un\beta}$ of the form (\ref{ref0}) 
\begin{eqnarray}
L^*_{\rho_{\un\beta}} = \Pi L\Pi+\sigma_{\un\beta}, \qquad {\rm with}  \quad
\sigma_{\un\beta}= - \sum_{i=1}^{N-1} (\beta_{i+1}-\beta_i) j_i \, , 
\label{eq: sigma 2}
\end{eqnarray}
provided the collection of inverse temperatures $\un \beta=(\beta_1,\ldots,\beta_N)$ is such that $\beta^{-1}_1=T_1$ and $\beta^{-1}_N=T_N$. 
Thus any $\sigma_{\un\beta}$ such that $\beta^{-1}_1=T_1$ and $\beta^{-1}_N=T_N$ will satisfy the Gallavotti-Cohen symmetry relation (see section \ref{subsec: fluctuation theorem}).
This was already observed in \cite{Eckent,rbt}.

\subsection{Asymmetric periodic chain}
\label{sec: asymmetric}

Building on a previous work \cite{Lefevere},  we introduce new dynamics by adding a mechanical force which creates a current through the system. On a periodic lattice, these dynamics lead to non-equilibrium systems with non-vanishing currents and in this sense, they are reminiscent of the asymmetric processes in lattice gas dynamics \cite{spohn}.

We start with an heuristic discussion before giving the definition of the dynamics.
If the heat baths are at different temperatures,  the dynamics (\ref{eqq}) is no longer reversible
with respect to the Gibbs measure or any ``local equilibrium" measure $\rho_{\un\beta}$.   
As observed  in (\ref{eq: sigma 2}), the function $\sigma$ which breaks the reversibility is a linear combination of the local energy currents.  
We show that adding an appropriate mechanical force  allows to modify at will the coefficients of this linear combination. In particular, the reversibility may be restored in (\ref{eqq}) by tuning the intensity of the additional mechanical force.   From the point of view of the generalized detailed balance, the action of the mechanical force is equivalent to the action of the local temperature gradient.

Let us first show that one may choose a (non-Hamiltonian) force which modifies the coefficients of the combination of local currents of energy in $\sigma$.  Take as generator of the dynamics,
\begin{eqnarray}
\label{eq: dynamique asymm}
L = L_S+L_H+L_{\un \gl} \, ,
\end{eqnarray}
where $L_H$ is the generator of  the Hamiltonian dynamics (\ref{eq: LH}), 
$L_S$ the generator of the two stochastic reservoirs (\ref{eqp2})
$$
L_S=- \gamma p_1 \frac{\partial}{\partial p_1} + \gamma T_1 \frac{\partial^2}{\partial p_1^2}
 - \gamma p_N \frac{\partial}{\partial p_N} + \gamma T_N \frac{\partial^2}{\partial p_N^2} \, ,
$$ 
and the contribution of the mechanical drift is given by the antisymmetric operator 
\qq
L_{\un \gl} =- \sum_{i=1}^N \hf T_i \big( \gl_{i-1}U'(q_{i-1}-q_i)+\gl_iU'(q_{i}-q_{i+1}) \big)\frac{\partial}{\partial p_i} \, ,
\label{eq: generator asym}
\qqq
To study the reversibility (\ref{eq: detail 1}) properties of this dynamics, we first compute, 
\begin{eqnarray*}
\rho^{-1}_{\un\beta}  L_{\un \gl}^T   \rho_{\un\beta} 
&=& - \sum_{i=1}^N \hf(\gl_{i-1}U'(q_{i-1}-q_i) + \gl_i U'(q_{i} - q_{i+1}))p_i\\
&=&  - \sum_{i=1}^{N-1} \gl_i\hf(p_i+p_{i+1})U'(q_{i} - q_{i+1}) =  \sum_{i=1}^{N-1} \gl_i j_i \, .
\end{eqnarray*}
Since the operator $L_S+L_H$ satisfies a generalized detailed balance relation wrt 
$\sigma_{\un \beta} = - \sum_{i=1}^{N-1} (\beta_{i+1}-\beta_i) j_i$  (\ref{eq: sigma 2}), we see that the dynamics (\ref{eq: dynamique asymm}) satisfies now the generalized detailed balance (see definition
\ref{def: generalized detailed balance})  for the measure  $\rho_{\un\beta}$ (\ref{ref0}) with
respect to $\sigma_{\un\beta, \un \gl}$
\begin{eqnarray}
L^*_{\rho_{\un\beta}} = \Pi L\Pi+\sigma_{\un\beta, \un \gl}, \qquad {\rm with}  \quad
\sigma_{\un\beta, \un \gl}=  \sum_{i=1}^{N-1} (\beta_i  - \beta_{i+1} + \gl_i) j_i \, .
\label{eq: sigma 3.0}
\end{eqnarray}
 When the relation $\gl_i\neq(\beta_{i+1}-\beta_i)$ for some $i$, then the system is driven out of equilibrium.
However, with the choice $\gl_0=\gl_N=0$ and $\gl_i=(\beta_{i+1}-\beta_i)$ reversibility is restored.  This is the key point which allows to identify  the strength of the mechanical force with the action of the local gradient of temperature.

For $\gl_i=(\beta_{i+1}-\beta_i)$ and slowly varying temperatures of the form $\beta_i^{-1} = T_i=T(\frac{i}{N})$ where $T$ is a smooth function, the generator $L_{\un \gl}$ becomes at lowest order in $\frac{1}{N}$,
\begin{eqnarray}
L_{\un \gl}=  \sum_{i=1}^N\frac{1}{N}\frac{\nabla T(\frac{i}{N})}{2 T(\frac{i}{N})}
\big( U'(q_{i-1} - q_i) + U' (q_{i}-q_{i+1}) \big) \frac{\partial}{\partial p_i} \, .
\label{eq: asym0}
\end{eqnarray}
Note that in the sum over $i$, the prefactors depending on the temperature and its gradient are basically constant when $i$ varies over distances much smaller than $N$.  

The asymmetric periodic chain \cite{Lefevere}, described below, is a chain with periodic boundary conditions and a dynamics made of three parts.  The first one corresponds to a usual Langevin dynamics for each oscillator, the second one to the Hamiltonian dynamics on the lattice and the third one is the previous generator (\ref{eq: asym0}) with constant prefactors (but arbitrary values). Namely, it is defined as 
\qq
L=L_S+L_H +L_\tau\,,
\qqq 
with,
\qq
L_S= \sum_i-\gamma p_i\frac{\partial}{\partial p_i}+\gamma T\frac{\partial^2}{\partial p^2_i} \, .
\qqq
 $L_H$ is defined in (\ref{eq: LH}) 
and the part driving the system out of equilibrium is,
\qq
L_\tau= - \frac{\tau}{2T}\sum_{i}(U' (q_{i-1}-q_{i})+ U' (q_i-q_{i+1}))\frac{\partial}{\partial p_i}\, .
\qqq

\noindent In terms of equations of motion the dynamics is described as follows,
\beqnarray
\label{dynamics}
dq_i&=&p_i dt \\
dp_i&=&-\gamma p_i dt-\frac{\partial H}{\partial q_i} dt
- \frac{\tau}{2T}(U'(q_{i-1}-q_i) + U' (q_i-q_{i+1}))dt
+ \sqrt{2\gamma T}dw_i \, , \nonumber
\eeqnarray
where $\tau \in {\bf R}$ is the new parameter regulating the strength of the non-equilibrium forcing and the $w_i$ are standard independent Brownian motion $i=1,\ldots,N$. Periodic boundary conditions means $q_0 = q_N$ and $q_{N+1} = q_1$.
Compared to the dynamics (\ref{eq: dynamics bulk}),  
the new term proportional to $\tau$ is the non-equilibrium part of the dynamics.
As  should be clear from the above argument it is responsible for the breaking of the time-reversal symmetry of the equilibrium dynamics ($\tau=0$).  Indeed, for the  dynamics (\ref{dynamics}), taking the Gibbs measure (\ref{eq: Gibbs}) at constant  temperature $T$ as the reference measure,
the generalized detailed balance holds  with a function $\sigma$ proportional to the total current (\ref{eq: current})
\begin{eqnarray}
\label{eq: sigma 3}
\sigma= \frac{\tau}{T^2} \sum_{i=1}^N j_i = \frac{\tau \, N}{T^2} J.
\end{eqnarray}
We will show in subsection \ref{positivity} that when $\tau\neq 0$, $\tau\langle J\rangle_{\hat\rho}>0$ for a stationary measure $\hat\rho$. Thus, the dynamics ensures the existence of an average non-vanishing energy current in the stationary state.
In section \ref{sec: Large deviations of the current}, 
the current large deviations for the forced dynamics (\ref{dynamics}) with harmonic potentials will be computed.
We will see there that (\ref{dynamics}) are the optimal dynamics to realize  current deviations.

\section{The large deviation functional}
\label{sec: DV}

The goal of this section is to rephrase the Donsker-Varadhan theory \cite{DonskerVaradhan,DonskerVaradhan2} in the framework of the 
Hamiltonian systems coupled to Gaussian stochastic thermostats and to discuss the relation with entropy production.


\subsection{The Donsker-Varadhan functional}
\label{subsec: DV functional}

In order to cover all the examples introduced in sections \ref{eq: Heat in the bulk}  and \ref{sec: asymmetric},
we consider the general dynamics defined by 
\beqnarray
\label{dynamics mixte 0}
dq_i&=&p_i dt \\
dp_i&=&-\gamma_i p_i dt-\frac{\partial H}{\partial q_i} dt
- \hf T_i(\gl_{i-1}U'(q_i-q_{i-1})+\gl_iU'(q_{i+1}-q_{i})) dt +\sqrt{2\gamma_i T_i}dw_i \, , \nonumber
\eeqnarray
with $\gamma_i >0, T_i > 0$ for $i= 1, \dots,N$. 
We stress the fact that the noise acts at each site $i$.
To simplify notation, we restrict to open systems  (with  the convention $q_0 = q_{N+1} = 0$,  $\gl_0 = \gl_N = 0$), but similar results hold with  periodic  boundary conditions (with the convention $q_{N+1} = q_1$ and $q_0 = q_N$). We denote by $P$ the probability of the evolution (\ref{dynamics mixte 0}) starting from a given 
initial data (which will play no role in the large $t$ asymptotic).

Typically, we shall be interested in the deviations over time of some physical quantities like the heat current (\ref{eq: current}).
Let the empirical distribution $\nu_t$ be defined by,
\begin{eqnarray*}
\nu_t (A)=\frac{1}{t} \int_0^t {\bf 1}_A \big( (\un p(s), \un q(s)) \big) ds \, ,
\label{empirical}
\end{eqnarray*}
where ${\bf 1}_A$ is the indicator function of a set $A\subset{\bf R}^{2N}$.  If the dynamics is ergodic $\nu_t$ converges to the ergodic invariant measure.  We now look at the asymptotic probability $P[ \nu_t \simeq \mu ]$ that the empirical distribution is 
close to the distribution of a given measure $\mu$ (in the sense of the weak convergence topology \cite{Wu})  for large $t$.  
Under suitable hypothesis (see Proposition \ref{prop: LD thme}), the dynamics obeys a large deviation principle with rate function $I$ and asymptotically in $t$
\qq
P[ \nu_t \simeq \mu ]\sim \exp \big( - t I(\mu) \big) \, .
\label{GDs}
\qqq
Furthermore, from the Donsker-Varadhan theory \cite{DonskerVaradhan, DonskerVaradhan2, Wu} the functional is given by 
\qq
I(\mu)=\sup_{g} \left\{ -\left< {L g \over g} \right>_\mu \; \Big | \quad g \in C_b^\infty ( {\bf R}^{2N}; [1, \infty[)
\right\} \, , 
\label{definition}
\qqq
where $C_b^\infty ( {\bf R}^{2N}; [1, \infty[)$  is the set of bounded infinitely differentiable functions in ${\bf R}^{2 N}$ taking values larger or equal to 1. We refer to Lemma 6.3.7 of \cite{DeuschelStroock} for the variational expression of $I$ in the case of hypoelliptic diffusions.

\medskip

The following Proposition justifies the validity of the large deviation principle (\ref{GDs}).
\begin{pro}
\label{prop: LD thme}
Suppose that the potentials $V$ and $U$ of the Hamiltonian (\ref{Hamilton}) are convex, twice differentiable and satisfy 
\begin{eqnarray}
V''(q) \ge \delta 
\quad {\rm and} \quad
\sum_{i=1}^N  V(q_i)+  {U(q_i-q_{i+1}) + U(q_{i+1} - q_i) \over 2}  \geq \delta \sum_{i=1}^N  \big( U' (q_i-q_{i+1}) \big)^2 \, , \nonumber\\
\label{eq: assump pot}
\end{eqnarray}
for some constant $\delta>0$. 

The dynamics (\ref{dynamics mixte 0}) (with $\forall i, \gamma_i>0$) obeys a large deviation principle with a functional $I$ given by the  Donsker-Varadhan theory  (\ref{definition})
provided $\max_i |\gl_i| \le \tau_0 $ and $\max_i |T_i - T_{i+1}| \le \Delta_0$, where $\Delta_0$ and $\tau_0$
are two constants  depending only on $\delta, \{\gamma_i,T_i\}_i$.
Furthermore,  the previous assumptions on the potentials (\ref{eq: assump pot}) ensure that the current (\ref{eq: current}) is exponentially integrable: for any $\lambda$ small enough
\begin{eqnarray}
\limsup_{t \to \infty} 
{1 \over t} \log P \left( \exp \left( \lambda \int_0^t \, ds  J (\un p(s), \un q (s)) \right) \right) < \infty \, .
\label{eq: tightness current}
\end{eqnarray}
\end{pro}

The proof heavily relies on previous results in the paper \cite{Wu} and it is postponed  to the Appendix.
The assumptions of Proposition \ref{prop: LD thme} on the potentials  $V$ and $U$ are not optimal, but they are sufficient to cover a wide class of physical examples.
In particular, similar statements hold also for any local modifications of the potentials  $V$ and $U$.
Remark that when the reservoirs act only at the boundary ($\gamma_i =0, i \in \{2,\ldots,N-1\}$), the large deviation principle for the current has been justified  in \cite{rbt2} (for a different class of Hamiltonians).

\bigskip

We are going to rewrite the functional $I$ (\ref{definition}) in a more explicit form.
The generator of the dynamics (\ref{dynamics mixte 0}) can be decomposed as $L=L_S+L_A$, 
with a symmetric part 
$L_S$ (\ref{eq: LS}) due to the noise and an antisymmetric part $L_A=\hf(L-\Pi L\Pi)$.
We introduce the notation,
\qq
\Gamma(f,g)=2\sum_{i=1}^N \gamma_i T_i \; (\partial_{p_i}f) \; (\partial_{p_i} g) \, , 
\qqq
for any smooth functions $f,g$.

\begin{pro}
\label{pro: 1}
Let $\rho$ be a measure and  $\sigma (\un q,\un p)$ a function such that 
\begin{eqnarray}
\label{eq: hyp adjoint}
L^*_{\rho} = \Pi L\Pi+\sigma \, . 
\end{eqnarray}
Let $\mu$ be a measure  absolutely continuous with respect to the measure $\rho$, with $f=d\mu/d\rho \in C^\infty ( {\bf R}^{2N}; {\bf R})$ such that $ \left< \| \nabla_{\un p}  \sqrt{f} \|_2 \right>_{\rho} < \infty$.
Then the functional $I$ (\ref{definition}) is given by
\qq
I(\mu)=-\left<f^\hf L_Sf^\hf\right>_{\rho}+ K(\mu)-\hf \left<\sigma\right>_{\mu} \, ,
\label{lemma1}
\qqq
with 
\qq
K(\mu)=-\inf_{W} \left( \frac{1}{8} \left<\Gamma(W,W)\right>_{\mu}+\hf \left <L_A W \right>_{\mu}\right)\geq 0 \, ,
\label{defK}
\qqq
where the infimum is taken over the smooth functions  $W \in C^\infty ( {\bf R}^{2N}; {\bf R})$ such that 
$W \in L^2(\mu)$ and $| \nabla_{\un p} W | \in L^2(\mu)$.
\end{pro}

From (\ref{eq: sigma 3.0}), we see that the dynamics (\ref{dynamics mixte 0}) satisfies the generalized detailed balance for the measure 
$\rho_{\un\beta}$ (\ref{ref0}), the assumption (\ref{eq: hyp adjoint}) of Proposition  \ref{pro: 1}  is satisfied  
with $\rho_{\un\beta}$ and $\sigma_{\un\beta, \un \gl}$
\begin{eqnarray*}
L^*_{\rho_{\un\beta}} = \Pi L\Pi+\sigma_{\un\beta, \un \gl}, \qquad {\rm with}  \quad
\sigma_{\un\beta, \un \gl}=  \sum_{i=1}^{N-1} (\beta_i  - \beta_{i+1} + \gl_i) j_i \, .
\end{eqnarray*}
Remark that the generalized detailed balance requires symmetry assumptions on the reference measure and on $\sigma$  which are not necessary for the Proposition \ref{pro: 1} to hold.


\bigskip

\noindent {\it Proof.}  Let $f=d\mu/d\rho$ and $g \in C_b^\infty ( {\bf R}^{2N}; [1, \infty[)$, then,
\qq
\left<g^{-1}L g\right>_{\mu}=\left<f g^{-1}L g\right>_{\rho}.
\label{ldf0}
\qqq
Write now in (\ref{ldf0}), $g=\sqrt{h f}$ for some $h>0$ and symmetrize the expression with the help of the adjoint operator of $L$,
\qq
\left<f g^{-1}L g\right>_{\rho}=\hf\left(\left<f^\hf h^{-\hf}L (h^\hf f^\hf)\right>_{\rho}+\left<f^\hf h^{\hf}L ^*(h^{-\hf} f^\hf)\right>_{\rho}\right)
\qqq
We note the relation,
\qq
L(\phi\psi)=\phi L\psi+\psi L\phi+\Gamma(\psi,\phi),
\qqq
And analogously for $L^*_\rho=\Pi L \Pi + \sigma = L_S-L_A+\sigma$ (by hypothesis),
\qq
L^*_\rho(\phi\psi)=\phi (L_S-L_A)\psi+\psi (L_S-L_A)\phi+\Gamma(\psi,\phi)+\sigma\phi\psi.
\qqq
Applying those formulas for $L$ (resp. $L^*_\rho$) with $\phi=h^\hf$ and $\psi=f^\hf$ (resp. $\phi=h^{-\hf}$ and $\psi=f^\hf$) using the 
generalized detailed balance relation $L^*_\rho = \Pi L \Pi + \sigma$ and computing systematically all derivatives, one gets,
\qq
\left<f g^{-1}L g\right>_{\rho}=\left<f^\hf L_S f^\hf\right>_{\rho}+\frac{1}{8}\left<\frac{\Gamma(h,h)}{h^2}\right>_{\mu}+\hf\left <h^{-1}L_Ah\right>_{\mu}+\hf\left<\sigma\right>_\mu.
\qqq
Writing $h=e^W$ and using the definition (\ref{definition}) of $I(\mu)$, one finally gets the
variational formula (\ref{defK}) restricted to functions $W$ such that 
$(W + \log f) \in C_b^\infty ( {\bf R}^{2N}; {\bf R})$.
To extend the class of functions $W$ in the variational formula (\ref{defK}), we consider an appropriate sequence of functions $\psi_n \in C^\infty ( {\bf R}^{2N}; {\bf R})$ such that 
$\psi_n = 1$ in the ball of radius $n$ and $\psi_n = 0$ outside the ball of radius $n +1$. Then the sequence 
$W_n = W \psi_n - (1 - \psi_n) \log f $ approximates any function $W \in C^\infty ( {\bf R}^{2N}; {\bf R}) \cap L^2 (\mu)$  with $| \nabla W| \in  L^2 (\mu)$.

The fact that $K(\mu)\geq 0$ follows by choosing $W=0$ in (\ref{defK}).

\vspace{5mm}

One of the important features of the representation (\ref{lemma1}) for $I(\mu)$ is the presence of $K(\mu)$. As we shall see below, in Proposition \ref{pro: K} and in the example (\ref{eq: exemple}), it is in general neither infinite or zero even in the case of dynamics satisfying detailed balance, i.e for equilibrium dynamics.  
$K(\mu)$ gathers the thermalizing effect of the noise which is transmitted from the $\un p$ variables to $\un q$ variables.  It corresponds to the ``traffic" in the terminology of \cite{MaesNetocny3}.   It is invariant under reversal of the sign of the momenta (i.e $K(\mu)=K(\Pi\mu)$) and when non-equilibrium forces are included it is invariant under a change of sign of the non-equilibrium parameter as we shall see in the gaussian systems of the last section. It measures the dynamical ``activity" in the Hamiltonian system irrespective of the sign of the non-equilibrium parameter.  
We shall see its role in the variational characterization of stationary states in section \ref{subsec: Entropy production}. We now compute a more explicit form for $K(\mu)$.

\vspace{5mm}

\begin{pro}
\label{pro: K}
Let $\mu$ be a measure on ${\bf R}^{2N}$ with smooth density wrt the Lebesgue measure $d \mu (\un p,\un q) = \exp( - \Phi ) d \un p d \un q$.
If  $\ov W \in L^2(\mu)$ is a solution of 
\qq
\sum_{i = 1}^N
- \gamma_i T_i \partial^2_{p_i} W +  \gamma_i T_i \;  \partial_{p_i} \Phi \partial_{p_i}W=-L_A \Phi \, ,
\label{HJ}
\qqq 
with $| \nabla_{\un p} \ov W | \in L^2(\mu)$ then 
\qq
K(\mu)=\frac{1}{8}\left<\Gamma(\ov W,\ov W)\right>_\mu=-\frac{1}{4}\left<L_A\ov W\right>_\mu \, .
\label{formK}
\qqq
Moreover, $\left<\Gamma(\ov W,\ov W)\right>_\mu$ is independent from the solution of (\ref{HJ}).
Finally, $K$ is symmetric wrt time reversal $K(\Pi \mu) = K(\mu)$.
\end{pro}

\noindent{\it Proof.}  Starting from the definition (\ref{defK}) of $K(\mu)$ and integrating by parts, leads to
\beqnarray
\frac{1}{8}\left<\Gamma(W,W)\right>_\mu=
\frac{1}{4} \sum_i  - \gamma_i T_i \left<W \partial^2_{p_i} W \right>_\mu
+ \gamma_i T_i  \left<W \partial_{p_i} \Phi \partial_{p_i} W \right>_\mu.
\label{intg}
\eeqnarray
Integrating also by parts the second term in (\ref{defK}) gives, 
\qq
\hf\left<L_AW\right>_\mu=\hf\left<W L_A \Phi \right>_\mu.
\label{intla}
\qqq
Therefore, the solution to the variational problem in (\ref{defK}) gives the equation (\ref{HJ}). Using then (\ref{intg}) and (\ref{intla}) and the definition (\ref{defK}) of $K(\mu)$ finally yields (\ref{formK}). 
The second equality in (\ref{formK}) follows by combining (\ref{intg}) and (\ref{HJ}).  To see that $K(\mu)$ does not depend on the solution to (\ref{HJ}), let $W,\tilde W$ be two different solutions, then,
\begin{eqnarray*}
\sum_{i = 1}^N - \gamma_i T_i  \partial_{p_i}^2 (W-\tilde W)+ \gamma_i T_i  \partial_{p_i}   \Phi  \partial_{p_i}  (W-\tilde W)=0.
\end{eqnarray*}
Multiplying this by $W+\tilde W$ and integrating by parts with respect to $\mu$, yields 
\begin{eqnarray*}
\left<\Gamma(W+\tilde W,W-\tilde W)\right>_\mu=0,
\end{eqnarray*}
and thus, $\left<\Gamma(W,W)\right>_\mu=\left<\Gamma(\tilde W,\tilde W)\right>_\mu$. 

\medskip

The measure $\Pi \mu$ has a density given by $\exp( - \Pi \Phi )$ and  $- \Pi \ov W$ is the corresponding solution of  (\ref{HJ}). Thus, the symmetry of $K$ follows from the identity 
\begin{eqnarray*}
K(\Pi \mu)=\frac{1}{8}\left<\Gamma(\Pi \ov W,\Pi \ov W)\right>_{\Pi \mu}
=\frac{1}{8}\left<\Gamma(\ov W,\ov W)\right>_\mu = K(\mu) \, .
\end{eqnarray*}
This concludes the proof.

\vspace{8mm}
Observe also that if $\mu$ is such that $L_A \Phi=0$ then $K(\mu)=0$. We want to emphasize that in general $K$ is non-zero 
and we will illustrate this in the simple case of a one dimensional harmonic oscillator coupled to a heat bath at temperature $T$
\qq
L=-\gamma p\frac{\partial}{\partial p}+\gamma T\frac{\partial^2}{\partial p^2}-\omega^2 q\frac{\partial}{\partial p}+p\frac{\partial}{\partial q} \, ,
\label{eq: exemple}
\qqq
with $L_S=-\gamma p\frac{\partial}{\partial p}+\gamma T\frac{\partial^2}{\partial p^2}$ and $L_A=-\omega^2 q\frac{\partial}{\partial p}+p\frac{\partial}{\partial q}$.
We will compute $K(\mu)$, where $\mu$ is the Gaussian measure  
$$
\mu(q,p)= {1 \over (2\pi b \omega^2)^{\hf} (2\pi a)^{\hf} }
\exp \left( -(a\frac{p^2}{2}+b\omega^2\frac{q^2}{2}) \right).
$$

In that case, (\ref{HJ}) becomes,
\begin{eqnarray*}
-\gamma T\partial_p^2W+\gamma Ta p\partial_p W=\omega^2 pq(b-a) \,.
\end{eqnarray*}
A solution is readily found,
\begin{eqnarray*}
\ov W=\frac{b-a}{a \gamma T}\omega^2 pq \, , 
\end{eqnarray*}
and then, from Proposition \ref{pro: K},
\qq
K(\mu)=\frac{1}{4\gamma T} \left( \frac{b-a}{a} \right)^2 \omega^4\left<q^2\right>_\mu=\frac{1}{4\gamma T} \left( \frac{b-a}{a} \right)^2
\frac{\omega^2}{b}.
\label{eq: K 1 oscillateur}
\qqq
From this example, we see that $K(\mu)$ vanishes only for Gaussian measures in which the average potential energy and kinetic energy are equal, i.e when equipartition of energy is realized.
In a sense, $K$ measures how much the Gaussian noise acting on the momenta variables $\un p$ is transmitted to the positions variables $\un q$.


\subsection{Entropy production} 
\label{subsec: Entropy production}

In this section, we show that the minimum entropy production principle does not apply for Hamiltonian dynamics and that the large deviation functional (\ref{lemma1}) is a natural extension for a variational characterization of the steady state.
Connections between large deviation functionals in stochastic systems and entropy production were initiated in \cite{MaesNetocny1}.
In order to identify the average entropy production of a dynamics in a given measure, we proceed as in \cite{spohn} in the context of interacting particle systems.  Namely, we define the entropy production as the difference between the variation of the Gibbs (Shannon) entropy and the transfer of heat by unit time due to the action of the thermostats.
A similar computation was performed in \cite{MaesNetocny0} for heat conduction networks and the identification of the average entropy production with the Dirichlet form of the process was obtained there and in \cite{Eckent}.

We consider  the dynamics (\ref{dynamics mixte 0}) for which the generator is given by $L = L_S + L_A$ with a symmetric part 
given by 
\begin{eqnarray*}
L_S=-\sum_{i=1}^N - \gamma_i p_i\frac{\partial}{\partial p_i}+ \gamma_i T_i\frac{\partial^2}{\partial p_i^2} \, .
\end{eqnarray*}
We first introduce the entropy production.
In thermodynamics, the entropy variation rate, or entropy production is the transfer of energy per unit time divided by the temperature at which the transfer takes place. Therefore, it is natural to define the average (with respect to a given measure $\mu$) entropy flux associated with the exchange of energy with the external heat baths as
\begin{eqnarray*}
\sigma_{ext}(\mu)=\left<\tilde L_SH\right>_\mu \, ,
\end{eqnarray*}
where 
\qq
\tilde L_S=\sum_{i=1}^N\gamma_i T_i^{-1}p_i\frac{\partial}{\partial p_i}+\sum_{i=1}^N\gamma_i \frac{\partial^2}{\partial p_i^2}.
\qqq
We have simply divided the contribution of each bath by its temperature $T_i$: when the temperatures are all equal to $T$ then 
$\tilde L_S= {1 \over T} L_S$. Computation yields,
\begin{eqnarray*}
\tilde L_S H=\sum_{i=1}^N\gamma_i(1-\frac{p_i^2}{T_i}) \, ,
\end{eqnarray*}
and thus the average entropy flux due to the coupling to the heat baths in the measure $\mu$ is,
\qq
\sigma_{ext}(\mu)=\sum_{i=1}^N\gamma_i\left<(1-\frac{p_i^2}{T_i})\right>_\mu.
\label{eq: sigma_ext}
\qqq
The Gibbs (or Shannon) entropy for any measure $\mu$ is,
\qq
S(\mu)=-\int dx f\log f.
\qqq
We define the entropy production in the chain as
\qq
s(\mu)\equiv\frac{d}{dt}S(\mu_t)|_{\mu_t=\mu}-\sigma_{ext}(\mu) \, .
\label{prodent}
\qqq
We show now that with this definition, the entropy production is always positive.
Note that,
\begin{eqnarray*}
\frac{d}{dt}S(\mu_t)=-\int dx f_tL\log f_t \, .
\end{eqnarray*}
As $L=L_S+L_A$ and $L_A$ is a first-order differential operator such that $L_A^\dagger=-L_A$, we see that
\begin{eqnarray*}
\frac{d}{dt}S(\mu_t)=-\int dx f_t(L_S+L_A)\log f_t=-\int dx f_t L_S\log f_t,
\label{derent}
\end{eqnarray*}
Comparing the density $f_t$ with the reference measure $\rho_{\un\beta}$ (see (\ref{ref0}))
\begin{eqnarray}
\rho_{\un\beta}={1 \over Z_{\un\beta}} \exp \left(- \sum_{i=1}^N \beta_i h_i \right) \, ,
\label{eq: rho}
\end{eqnarray}
where $\beta_i=T_i^{-1}$ for every $i$ such that $\gamma_i\neq 0$ (other $\beta_i$ are arbitrary), we get,
\begin{eqnarray}
\frac{d}{dt}S(\mu_t)&=&-\int dx f_t L_S\log \frac{f_t}{\rho_{\un \beta}}
+ \int dx f_t L_S \left( \sum_{i=1}^N \beta_i h_i \right) \nonumber \\
&=&-\int dx f_t L_S\log \frac{f_t}{\rho_{\un\beta}}+\sum_{i=1}^N\gamma_i \int f_t (1-\frac{p^2_i}{T_i}).
\label{eq: prod 0}
\end{eqnarray}
We recall that $\Gamma(f,g)  = 2 \sum_{i=1}^N\gamma_i T_i  \;  \partial_{p_i}f\partial_{p_i}g$.
For the first term, we note the identity,
\begin{eqnarray*}
L_S(\log h)=h^{-1}L_S h-\hf h^{-2}\Gamma(h,h) \, .
\end{eqnarray*}
Combining (\ref{eq: sigma_ext}), (\ref{eq: prod 0}) 
and using the fact that $(L_S)^\dagger \rho_{\un \beta}=0$ (because $\beta_i=T_i^{-1}$ for every $i$ such that $\gamma_i\neq 0$), with  (\ref{prodent}), this yields
\qq
s(\mu)=\hf \int dx \rho_{\un\beta}(\frac{\rho_{\un\beta}}{f})
\Gamma \left(\frac{f}{\rho_{\un\beta}},\frac{f}{\rho_{\un\beta}} \right)
= 2 \int dx \rho_{\un\beta} \Gamma \left( \sqrt{ \frac{f}{\rho_{\un\beta}}} \; , \; \sqrt{\frac{f}{\rho_{\un\beta}}} \right) \, ,
\label{entroprod}
\qqq
with $f$ the density of the measure $\mu$.
Since $\Gamma(f,f)\geq 0$, it is easy to see that $s(\mu)\geq 0$ and that the infimum is reached when $f=\rho_{\un \beta}$.  As already observed in \cite{MaesNetocny0},  it is also obvious that, as $\Gamma$ only involves derivatives with respect to the variables $\un p$, one can add any function of $\un q$ in the exponential defining $\rho_{\un\beta}$.  Therefore, even in the case of an equilibrium dynamics, when $T_i=T$, $\forall i$, such that $\gamma_i\neq 0$ in $L_S$, the minimum entropy production principle does not single out the equilibrium measure $\rho_{\un\beta}$.  
This comes from (\ref{derent}) which expresses the fact that the Gibbs entropy is invariant under the Hamiltonian evolution.  Entropy is produced solely by the action of the thermostats which act only on the $\un p$ variables.

\bigskip

As observed in \cite{MaesNetocny1}, the large deviation functional $I$ provides a natural 
variational characterization of the stationary measure, which as we will see below, generalizes the minimum production entropy principle. 
As far as the variational principle is concerned, the key observation is that $I(\mu)\geq 0$ and  $I(\mu)=0$ if and only if $\mu$ is a stationary measure for the process associated to the generator $L$ (Theorem 4.2.39 of \cite{DeuschelStroock}).
We apply now Proposition \ref{pro: 1} to the dynamics (\ref{dynamics mixte 0}) which satisfies a generalized detailed balance
(\ref{eq: sigma 3.0}) wrt  the reference measure $\rho= \rho_{\un\beta}$ (\ref{eq: rho}) and $\sigma_{\un\beta, \un \gl}=  \sum_{i=1}^{N-1} (\beta_i  - \beta_{i+1} + \gl_i) j_i$. Thus (\ref{lemma1}) reads
\qq
I(\mu)=\frac{1}{4} s(\mu) + K(\mu) - \hf\left<\sigma_{\un \beta,\un \gl}\right>_\mu \, ,
\label{dvent}
\qqq
 where we have used the following identity obtained by integration by parts from (\ref{entroprod})
\qq
\frac{1}{4}s(\mu)=-\left<f^\hf L_Sf^\hf\right>_{\rho_{\un\beta}} \, ,
\label{eq: ent prod}
\qqq
for the reference measure $\rho_{\un\beta}$.  
The first term in (\ref{dvent})  is identified with the entropy production $s$ of the measure $\mu$,  and as we have seen in the previous section, the second one $K$ records the coupling between the positions and momenta.
The last term in (\ref{dvent}) is the time-reversal breaking term given by combination of the microscopic currents.
For equilibrium dynamics, by definition, the term $\sigma$ is absent, and the presence of $K$ ensures that the minimization of the sum of the first two terms in (\ref{dvent}) yields the stationary measure univocally.   Indeed, as we have explained above, minimizing entropy production alone is not sufficient to determine the equilibrium distribution. The example (\ref{eq: exemple}) of a single harmonic oscillator coupled to a heat bath is interesting in that respect and we come back briefly to it. 
We have already computed $K(\mu)$ in (\ref{eq: K 1 oscillateur}) and compute now the entropy production $s(\mu)$.
We consider  the reference measure,
\qq
\rho (p,q) ={1 \over Z} \exp \left( -\frac{1}{T}(\frac{p^2}{2}+\omega^2\frac{q^2}{2}) \right) \, ,
\qqq
and a Gaussian measure 
$$
\mu(q,p)= {1 \over (2\pi b\omega^2)^{\hf}(2\pi a)^{\hf}}
\exp \left( -(a\frac{p^2}{2}+b\omega^2\frac{q^2}{2}) \right) .
$$
Then $s(\mu)=a^{-1}\gamma T(a-\beta)^2$. Combining this with the expression of $K(\mu)$ 
(\ref{eq: K 1 oscillateur}), we get,
\qq
I(\mu)=\frac{1}{4a}\gamma T(a-\beta)^2+\frac{1}{4 \gamma T} \left( \frac{b-a}{a } \right)^2  \frac{\omega^2}{b}.
\qqq
In this simple example, we see that the role of $K$ is to ensure the equipartion of energy by the action of the Hamiltonian dynamics, the role of the entropy production is to fix the temperature of the equilibrium distribution.

\subsection{Positivity of the energy current in the asymmetric chain.}
\label{positivity}

We come back to the asymmetric chain defined in (\ref{dynamics}) and use the previous results on the representation of the Donsker-Varadhan functional to prove that when the parameter $\tau$ is different from zero then in the stationary state $\hat \rho$, $\tau\langle  J\rangle_{\hat\rho}>0$. 
We note that the hypoellipticity of the process defined by the asymmetric chain may be shown by checking the H\"ormander condition on the generator as in \cite{rbt}.  For the asymmetric periodic chain, no coupling between nearest-neighbour  is even required because the noise acts on every particle.   Hypoellipticity of the process implies the smoothness of the probability transition and thus that the stationary state, whenever it exists, is described by a smooth density.  Irreducibility properties of the process may be checked by a control argument and Stroock-Varadhan  support theorem \cite{StroockVaradhan} as explained in \cite{Carmona,rbt}.  Hypoellipticity and irreducibility imply together that there is at most one stationary measure.
The existence of a (unique) stationary measure follows from those two properties and the existence of a Lyapunov function as in (\ref{eq: Phi}) in the appendix (see for instance Theorem  8.7 of \cite{rbln}).

\begin{pro}
Let $U$ and $V$ be potentials satisfying the hypothesis of proposition \ref{prop: LD thme} and such that $U''$ is not uniformly equal to 0. Let $\hat\rho$ be the unique stationary state for the dynamics (\ref{eq: dynamique asymm}), then $\tau\langle J\rangle_{\hat \rho}\geq 0$.  If $\tau\neq 0$, then  $\tau\langle J\rangle _{\hat\rho}> 0$.
\end{pro}

\noindent {\it Proof. }  
Take as a reference measure $\rho_\beta=Z^{-1}\exp (- \beta H)$, it was noted in (\ref{eq: sigma 3}) that the dynamics satisfies a generalized detailed balance relation with respect to $\rho_\beta$ and $\sigma$, with $\sigma=\frac{\tau N}{T^2}J$, $T=\beta^{-1}$.  Write the formula (\ref{lemma1}) of Proposition \ref{pro: 1} under the form (\ref{dvent}),
\qq
I(\mu)=\frac{1}{4}s(\mu)+K(\mu)-\hf \langle \sigma \rangle_\mu \, ,
\label{imuasym}
\qqq
for a smooth measure $\mu$. 
As $\hat\rho$ is a smooth stationary measure, then $I(\hat\rho)=0$.  Since $s(\hat\rho)$ and $K(\hat\rho)$ are positive,  one concludes that  $\tau\langle  J\rangle_{\hat\rho}\geq0$, using (\ref{imuasym}).  To show that if $\tau\neq 0$,  $\langle  J\rangle_{\hat\rho}>0$, we proceed by contradiction. Assume then that $\tau\neq 0$ and $\langle  J\rangle_{\hat\rho}=0$, then simultaneously $s(\hat\rho)=0$ and $K(\hat\rho)=0$, since the two are always non-negative.
Writing $\hat\rho=\rho_\beta\exp (-\varphi)$, the first condition implies by construction of $s(\hat\rho)$, (see (\ref{entroprod})) that, 
\qq
\partial_{p_i}\varphi=0\,,\;\forall i.
\label{cst0}
\qqq
On the other hand, $K(\hat\rho)=0$ implies that the solution $\ov W$ in (\ref{formK}) is such that $\partial_{p_i}\ov W=0\,,\;\forall i$.  Therefore, using (\ref{HJ}), $\varphi$ must be such that,
\qq
L_A(\beta H+\varphi)=0
\label{cst1}
\qqq
Since $L_A=L_H+L_\tau$ and combining  (\ref{eq: dynamique asymm}) with (\ref{cst0}), $\varphi$ satisfies
\qq
\sum_{i=1}^Np_i\partial_{q_i}\varphi=\frac{\tau}{2T}\sum_{i=1}^Np_i\Psi_i
\qqq
with $\Psi_i=U'(q_{i-1}-q_i)+U'(q_i-q_{i+1})$.  As this equation holds for any vector $\un p$ and because by (\ref{cst0}), $\varphi$ does not depend on $\un p$, it means that 
\qq
\partial_{q_i}\varphi=\frac{\tau}{2T}\Psi_i\,,\;\forall i.
\label{cst2}
\qqq
But since $\partial_{q_i}\Psi_{i+1}=U''(q_i-q_{i+1})$ and $\partial_{q_{i+1}}\Psi_{i}=-U''(q_i-q_{i+1})$, there cannot be any solution to (\ref{cst2}), unless $U''(q)=0$ for all $q$.  Therefore, we conclude that $\langle J\rangle_{\hat\rho}\neq0$ and thus $\tau\langle J\rangle_{\hat\rho}>0$, because  $\tau\langle J\rangle_{\hat\rho}\geq0$.

\subsection{The Gallavotti-Cohen symmetry}
\label{subsec: fluctuation theorem}

An interesting aspect of (\ref{lemma1}) in Proposition \ref{pro: 1},
\qq
I(\mu)=-\left<f^\hf L_Sf^\hf\right>_{\rho}+ K(\mu)-\hf \left<\sigma\right>_{\mu} \, ,
\label{basici}
\qqq
 is that when the triple $(L,\rho,\sigma)$  satisfies a generalized detailed balance relation, the three terms are either odd or even under the reversal of momenta $\Pi$.  The fact that $K(\Pi\mu)=K(\mu)$ was established in Proposition \ref{pro: K}.  In the previous section, we have identified the first term in (\ref{basici}) to the entropy production $s(\mu)$.  It  is also invariant under time-reversal (i.e  $s(\Pi\mu)=s(\mu)$) when $\Pi\rho=\rho$ because by construction $\Pi L_S\Pi=L_S$.  And, since $\sigma$ is odd under time reversal ($\Pi\sigma=-\sigma$), one may therefore write, 
 \qq
I(\Pi\mu)=-\left< f^\hf L_S f^\hf\right>_{\rho}+K(\mu)+\hf\left<\sigma\right>_{\mu} 
= I(\mu) +  \left<\sigma\right>_{\mu}.
\label{pinu}
\qqq
 Heuristically, (\ref{pinu}) implies the  Gallavotti-Cohen symmetry for the large deviations of the function $\sigma$ integrated over time
 \begin{eqnarray*}
\lim_{t \to \infty} \; - {1 \over t} \log P\left[\frac{1}{t}\int_0^t \; \sigma(s) ds=\hat\sigma \right] = \hat I(\hat\sigma) \, ,
\end{eqnarray*}
with 
\qq
\hat I(\hat\sigma)=\inf_\mu\{I(\mu)|\left<\sigma\right>_\mu=\hat\sigma\} \, .
\qqq
Since $\Pi\sigma=-\sigma$,
\begin{eqnarray*}
\hat I(-\hat\sigma)&=&\inf_\mu\{I(\mu)|\left<\sigma\right>_\mu=-\hat\sigma\} =\inf_\mu\{I(\mu)|\left<\Pi \sigma\right>_\mu=\hat\sigma\} \nonumber \\
&=&\inf_\mu\{I(\mu)|\left< \sigma\right>_{\Pi\mu}=\hat\sigma\} = \inf_\mu\{I(\Pi \mu)|\left< \sigma\right>_\mu=\hat\sigma\}\,.
\label{symm}
\end{eqnarray*} 
Thus  one gets from (\ref{pinu})
\qq
\hat I(\hat\sigma)-\hat I(-\hat\sigma)=- \hat\sigma.
\qqq
This is the celebrated Gallavotti-Cohen symmetry relation \cite{GallavottiCohen1,GallavottiCohen2,EvansCohenMorriss,Kurchan, Kurchan2,LebowitzSpohn, Maes}.
For a chain of oscillators in contact with two heat baths, the Gallavotti-Cohen symmetry has been rigorously justified in \cite{Eckent, rbt2}.

\section{Large deviations of the current}
\label{sec: Large deviations of the current}

Using the Donsker-Varadhan variational principle (\ref{lemma1}),  we  are going derive explicit expressions for the large deviation function in the case of harmonic interaction when the system size diverges. We consider the asymmetric dynamics (\ref{dynamics}) on a 
periodic chain of length $N$ where each oscillator is coupled to a heat bath at  temperature $T$
\beqnarray
\label{eq: gauss dynamics}
dq_i&=&p_i dt \\
dp_i&=&-\gamma p_i dt-\frac{\partial H}{\partial q_i} dt+\frac{\tau \omega^2}{2T} \big( q_{i+1} - q_{i-1} \big) dt
+\sqrt{2\gamma T} dw_i ,  \nonumber
\eeqnarray
with harmonic potential 
\beqnarray
H (\un q,\un p) =  \sum_{i = 1}^N   \frac{p_i^2}{2} + \frac{\omega_0^2}{2}q_i^2+ \frac{\omega^2}{2} (q_i-q_{i+1})^2 \, ,
\label{eq: gauss potential}
\eeqnarray
and a pinning $\omega_0 > 0$.\\

We will study the large deviations of the spatial average of the total  current (\ref{eq: current})
\begin{eqnarray}
\label{eq: current gauss}
J (\un q,\un p) = {\omega^2 \over 2 N} \sum_{i = 1}^N   (q_{i+1} - q_{i})(p_i+p_{i+1}).
\end{eqnarray}
The large deviation principle (Proposition \ref{prop: LD thme}) implies that for $\lambda$ and $\tau$ small enough 
\begin{eqnarray}
\label{eq: LD current}
\CF^\tau_{\lambda,N} = \lim_{t \to \infty} {1 \over t} \log P \left( \exp \left( N \lambda \int_0^t \; ds \; J (\un q(s),\un p(s))  \right) \right) \, ,
\end{eqnarray}
where 
\qq
\CF^\tau_{\lambda,N} = 
\sup_\mu\{\CF^\tau_{\lambda,N} (\mu)\}  \;\;\;{\rm with}\;\;\;
\CF^\tau_{\lambda,N} (\mu)= N\lambda\left<J\right>_\mu  - I^\tau(\mu) \, ,
\label{inf}
\qqq
where the upper-script $\tau$ emphasizes the dependency on the asymmetry.
We recall from Proposition \ref{pro: 1}  that  $I^\tau(\mu)$ can be written,
\qq
I^\tau(\mu)=\frac{1}{4}s(\mu)+K^\tau(\mu)-\frac{\tau N}{2T^2}\left< J\right>_\mu \, ,
\label{itau}
\qqq
where $s(\mu)$ denotes the entropy production (\ref{eq: ent prod}).

\bigskip

For the harmonic chain (\ref{eq: gauss dynamics}), Gaussian computations lead to the following asymptotics for $\CF^\tau_{\lambda,N}$
\begin{pro}
\label{pro: gaussian}
\ 

\noindent
(i) 
For $\tau =0$ and fixed $N$, one has for $\lambda$ small
\begin{eqnarray}
\label{eq: asymptotic tau=0}
\CF^{0}_{\lambda,N} = \sum_{k = 1}^{N} 
\frac{T^2  \omega^4} { \gamma \omega_k^2}  \sin \left( 2 \pi {k \over N} \right)^2 \lambda^2 + (  \frac{T^4 \omega^8} { \gamma \omega_k^6}  + 5 \frac{T^4 \omega^8} { \gamma^3 \omega_k^4} ) \sin \left( 2 \pi {k \over N} \right)^4 \lambda^4 
+ O \big( \lambda^6 \big) \, , \nonumber \\
\end{eqnarray}
with $\omega_k^2 = \omega^2_0  + 4 \omega^2 \sin \left( 2 \pi {k \over N} \right)^2$.

\noindent
(ii) Fix $\omega_0 > 0$ and two parameters $\lambda', \tau' \in {\bf R}$.
For large $N$, with the scaling $\lambda = \lambda' / N$  and $\tau = \tau ' / N$,  one gets
\qq
\lim_{N \to \infty} \; N \CF^\tau_{\lambda,N} = \kappa \big( \lambda' \tau' +  (\lambda')^2 T^2 \big) \, ,
\label{eq: CF N large}
\qqq
where $\kappa$ is the conductivity of the model  
\qq
\kappa=\frac{1}{\gamma}\int_{-\frac{1}{2}}^{\frac{1}{2}}\frac{\omega^4 \sin^2(2\pi x)}{\omega^2_0 
+ 4 \omega^2 \sin^2(2 \pi x)} \, dx \, .
\label{eq: kappa}
\qqq
\end{pro}

Before deriving Proposition \ref{pro: gaussian}, we first comment on the results. 
Taking the Legendre transform of (\ref{eq: CF N large}) shows that for a weak drift $\tau = {\tau' \over N}$, the large deviations of the currrent 
are given by
\begin{eqnarray*}
\lim_{t \to \infty} - {1 \over t} \log P \left(  \int_0^t \; ds \; J (\un q(s),\un p(s)) \simeq  {j \over N} \right)
= \hat I_N (j)
\end{eqnarray*}
where $ \hat I_N$ scales,  for large $N$, like
\qq
\label{eq: LD current 2}
\hat I_N  (j)= {1 \over N} \frac{(j- \kappa \tau' )^2}{4 \kappa T^2} + O \left( {1 \over N^2} \right) \, .
\qqq
In the stationary regime, the mean current has been already computed in terms of the  conductivity (\ref{eq: kappa}) (see \cite{Lefevere})
\begin{eqnarray*}
\lim_{N \to \infty}  \lim_{t \to \infty}  N \; \left<  J (\un q(t),\un p(t)) \right> = \kappa \tau' \, .
\end{eqnarray*}
The expression (\ref{eq: LD current 2}) is new in the framework of thermostated Hamiltonian systems:
it relates the linear response theory \cite{Lefevere} to the 
large deviations and shows that the atypical behavior can still be described by the conductivity $\kappa$. The reason is that the local equilibrium is preserved for small shifts of the current
(remark that the fourth order term in (\ref{eq: asymptotic tau=0}) corresponds to corrections to local
equilibrium).
For weak current deviations,  (\ref{eq: LD current 2}) has a Gaussian structure similar to the one obtained in lattice gases with a weak asymmetry 
\cite{BD2,jona2}. The variance in (\ref{eq: LD current 2}) (given by $2 \kappa T^2$) is related to the response coefficient $\kappa$ according to the usual Einstein fluctuation-dissipation.
In the derivation of (\ref{eq: asymptotic tau=0}), we will see that the corrections due to $K^{\tau =0}$ arise only at the order $\lambda^4$. 
The functional is only quadratic asymptotically in $N$. 
The Gaussian structure will not remain for non vanishing drifts $\tau$ or larger current deviations. 
A similar behavior can be observed in lattice gases like the weakly/totally asymmetric exclusion processes \cite{BD2}.

\bigskip

\noindent
{\it Proof of Proposition \ref{pro: gaussian}}.

As the system is periodic and the functional $\CF^\tau_{\lambda,N}$ is convex,
one can look for a maximizer of the variational principle (\ref{inf}) among the translation-invariant measures.
The current $J$ in (\ref{eq: current}) is quadratic and the evolution equations (\ref{eq: gauss dynamics}) linear, thus the minimum in (\ref{inf}) will be reached for Gaussian measures (to see this, one can check that the maximizer of  (\ref{inf}) is obtained in terms of the principle eigenvector of the operator $L-\lambda J$ and the principle eigenvector of its adjoint).
Thus it remains to optimize the variational principle (\ref{inf}) over Gaussian translation-invariant measures.
We state first Gaussian estimates before  deriving (i) and (ii).

\bigskip

\noindent
{\bf Gaussian estimates:}

For periodic systems, the  Gaussian computations are easier in Fourier coordinates.
We set
\qq
P_k=\frac{1}{\sqrt{N}}\sum_{j=1}^N e^{i\frac{2\pi}{N}k j} p_j,
\qquad
Q_k=\frac{1}{\sqrt{N}}\sum_{j=1}^N e^{i\frac{2\pi}{N}k j}q_j \, .
\nonumber
\qqq
For notational simplicity, we choose an odd length $N = 2 n +1$ and the Fourier modes $k \in \{ -n,n \}$.
This choice leads to symmetric relations: $\ov P_k=P_{-k}$ and thus, $|P_k|^2=P_{-k}P_k$ and similarly  for the variables $Q_k$. 

\bigskip

We will first provide explicit expressions of $I^\tau(\mu)$ for  a general Gaussian measure $\mu$.
Given the reference measure $\rho_0$ 
\qq
\rho_0=\exp \left(- {1 \over 2 T} \sum_k  |P_k|^2 + \omega_k^2 \; |Q_k|^2   \right) \, ,
\qqq
the most general form for a Gaussian translation-invariant measure in the variables $(\un P,\un Q)$ may be written  $\mu= Z^{-1}\exp(-\frac{1}{T} \Psi) \rho_0$, with
\begin{eqnarray}
\Psi= \sum_{k = - n}^{n} \left( \ii \; a_k P_k Q_{-k} +
 b_k \big[ |P_k|^2 + \omega_k^2 \; |Q_k|^2  \big] +c_k \big[ |P_k|^2 - \omega_k^2 \; |Q_k|^2   \big] \right)
 \label{psi}
\end{eqnarray}
with $a_k =  \Re (a_k) + \ii  \Im(a_k) = - \ov a_{-k} \in \NC$ and $b_k = b_{-k} \in \NR$, $c_k = c_{-k} \in \NR$. 
The expectation under $\mu$ will be denoted $ <  \cdot >_\mu$, and for each mode $k \in \{-n,n \}$
\begin{eqnarray}
\label{covariance}
&&  \left <  |P_k|^2 \right >_\mu =\delta_k(1 + 2 (b_k-c_k)) \omega_k^2 T
, \quad  
\left<  |Q_k|^2   \right >_\mu = \delta_k (1 + 2 (b_k+c_k)) T\\
&& {\rm i} \left<  P_k Q_{-k}  - P_{-k} Q_k   \right >_\mu = - 2  \Re (a_k) \delta_k T,
\quad 
\left<P_k Q_{-k}  + P_{-k} Q_k   \right >_\mu=-2 \Im(a_k) \delta_k T\nonumber 
\end{eqnarray}
with $\delta_k^{-1}=(1 + 2 b_k)^2 \omega_k^2-4c_k^2\omega_k^2 - |a_k|^2$.

\bigskip

Using now Proposition \ref{pro: K}, we are going to compute $K^\tau(\mu)$ for the dynamics and $\mu$ as above.
In Fourier coordinates the antisymmetric part of the generator $L_A =L_H+L_{\tau}$ (\ref{generator}) is given by
\qq
L_H=\sum_{k = -n}^n P_k\frac{\partial}{\partial Q_k}-\omega^2_kQ_k\frac{\partial}{\partial P_k}
\label{LHgaussien}
\qqq
with $\omega_k^2 = \omega^2_0  + 4 \omega^2 \sin \left( 2 \pi {k \over N} \right)^2$ and 
\qq
L_\tau=-\ii \frac{\tau\omega^2}{T} \sum_{k = -n}^n  \sin(\frac{2\pi k}{N})Q_k\frac{\partial}{\partial P_k} \, .
\label{Ltaugaussien}
\qqq

We first show that if $\Im (a_k)\neq 0$ for some $k$, then $K^\tau (\mu)=+\infty$.   
Remark that $L_A |Q_k|^2= P_k Q_{-k}  + P_{-k} Q_k $ and that according to (\ref{covariance}), $\left< L_A |Q_k|^2 \right >_\mu=0$ if and only if   $\Im (a_k)= 0$.  
From the definition of $K^\tau (\mu)$ (\ref{defK}),  if  $\Im (a_k)\neq 0$ for some $k$, it is possible to choose a sequence of real numbers $\zeta_n$ such that for $W_n= \zeta_n |Q_k|^2$
$$
\lim_{n\rightarrow\infty}\left(\frac{1}{8}\left<\Gamma(W_n,W_n)\right>_{\mu}+\hf \left <L_A W_n \right>_{\mu}\right)
=\lim_{n\rightarrow\infty} \hf \zeta_n\left<P_k Q_{-k}  + P_{-k} Q_k   \right >_\mu=-\infty \, ,
$$ 
and thus $K^\tau (\mu)=+\infty$.
So from now on, we assume that $a_k$ is real and that $a_k=-a_{-k}$.
In that case a solution of (\ref{HJ}) is
\qq
\ov W=
\sum_{k = - n}^n   \sin \left( 2\pi\frac{k}{N} \right)\frac{\ii \tau\omega^2}{2\gamma T^2} \big( P_kQ_{-k}-P_{-k}Q_k \big)
+\frac{ c_k\omega_k^2}{(\hf+b_k+c_k)\gamma T}(P_kQ_{-k}+P_{-k}Q_k).
\qqq
By Proposition \ref{pro: K}, one gets,
\qq
\label{eq: K exact}
K^\tau(\mu)=\frac{1}{\gamma} \sum_{k = - n}^n  
\delta_k(1+2(b_k+c_k)) \left( \frac{\tau^2\omega^4}{4T^2} \sin \left( 2\pi\frac{k}{N} \right)^2
+\frac{4c_k^2\omega_k^4}{(1+2(b_k+c_k))^2} \right) \, ,
\qqq
with $\omega_k^2 = \omega^2_0  + 4 \omega^2 \sin \left( 2 \pi {k \over N} \right)^2$
and $\delta_k^{-1}=(1 + 2 b_k)^2 \omega_k^2-4c_k^2\omega_k^2 - a_k^2$.

Note that $K^\tau(\mu)$  is made of two parts: an equilibrium one (independent of the forcing $\tau$) and of a non-equilibrium one.

\bigskip

\noindent The entropy production (\ref{entroprod}) is given by
\begin{eqnarray*}
&& s(\mu) = \frac{\gamma}{T} \sum_{k = - n}^{n}  \left< \frac{\partial \Psi}{\partial P_k} \frac{\partial \Psi}{\partial P_{-k}}  \right>_\mu \\
&& \qquad = \frac{\gamma}{T}
\sum_k 2 \ii   a_k(b_k+c_k) \left<P_kQ_{-k}-P_{-k}Q_k\right>_\mu
+ a^2_k\left<|Q_k|^2\right>_\mu+4 (b_k+c_k)^2\left<|P_k|^2\right>_\mu\,.
\end{eqnarray*}
For $a_k$ real, one gets from (\ref{covariance})
\beqnarray
s(\mu)=  \gamma 
\sum_{k = - n}^{n}  \;    \delta_k \Big[ a^2_k(1 - 2 (b_k+c_k))+4 (b_k+c_k)^2\omega^2_k(1+2(b_k-c_k)) \Big] \, .
\eeqnarray

\bigskip

\noindent Finally, the current can be represented in Fourier coordinates
\begin{eqnarray*}
J = - \ii \; \frac{\omega^2}{2 N} \sum_{k = - n}^n 
\sin \left(2 \pi \frac{k}{N} \right) \left( P_k Q_{-k} -  P_{-k} Q_k \right) \, ,
\end{eqnarray*}
and from (\ref{covariance}), its expectation is 
\begin{eqnarray}
<J>_\mu  =  \frac{\omega^2 T}{N} \sum_{k = - n}^n 
\sin \left(2 \pi \frac{k}{N} \right)   a_k \delta_k \, .
\label{eq: mean current}
\end{eqnarray}

\bigskip

\noindent
{\bf Proof of  (i) :}

\noindent The functional (\ref{inf}) reads for $\tau = 0$
\begin{eqnarray*}
\label{eqfunc}
\CF^\tau_{\lambda,N} (\mu)= - \frac{1}{4}s(\mu) - K^0 (\mu) + N\lambda\left<J\right>_\mu\,.
\end{eqnarray*}
Restricting to translation invariant Gaussian measures (\ref{psi}), the previous Gaussian computation lead to 
\beqnarray
&& \CF_{\lambda,N} (\mu)= 
\sum_{k = -n}^n \delta_k \left[ \lambda T  \omega^2 \sin \left(2 \pi \frac{k}{N} \right) a_k -
\frac{1}{\gamma}  \frac{4 c_k^2 \omega_k^4}{(1+2(b_k+c_k))} \right. \\
&& \qquad \qquad \qquad 
 - \left. \frac{\gamma}{4}\left(a^2_k(1 - 2(b_k+c_k))+4 (b_k+c_k)^2 \omega_k^2 (1+2(b_k-c_k)) \right) \right] \nonumber
\eeqnarray

Optimizing over each Fourier mode leads at the order $\lambda^2$
\begin{eqnarray}
&& a_k =  {2 T  \omega^2 \over \gamma}  \sin \left(2 \pi \frac{k}{N} \right)   \lambda + O (\lambda^3), 
\quad
b_k = - \left( {T^2 \omega^4 \over 2 \omega_k^4} + {T^2 \omega^4 \over \gamma^2 \omega_k^2} \right)   \sin \left(2 \pi \frac{k}{N} \right)^2 \lambda^2,
\nonumber \\
&& c_k =  {T^2  \omega^4 \over 2 \omega_k^4}  \sin \left(2 \pi \frac{k}{N} \right)^2  \lambda^2 \, . 
\label{eq: coeff}
\end{eqnarray}
The previous computations imply that at the order $\lambda$ the solution of the variational principle  (\ref{inf}) is given by
\qq
\mu (\un q,\un p)=
{1 \over Z} \exp\left( \frac{2\lambda N}{\gamma}  J(\un q, \un p)\right) \rho_0 (\un q,\un p) \, .
\label{harmeasure0}
\qqq
Finally, asymptotics  (\ref{eq: asymptotic tau=0}) follow from (\ref{eq: coeff}).

\vskip1cm

\noindent
{\bf Proof of (ii) :}

\noindent We observe that (\ref{inf}) may be rewritten
\qq
\sup_\mu \CF^\tau_{\lambda,N} (\mu) =
\sup_\mu \left[- I^{\tau+2\lambda T^2}(\mu) - K^\tau(\mu) + K^{\tau+2\lambda T^2}(\mu) \right].
\label{itau2}
\qqq
The point of writing  the variational problem under this form is that minimizing only $ I^{\tau+2\lambda T^2}$ is easy.  The minimizer is the stationary measure of the asymmetric periodic chain with non-equilibrium parameter $\tau$ replaced by $\tau+2\lambda T^2$.  It was derived in \cite{Lefevere},
\qq
\rho_{\tau+2\lambda T^2}(\un q,\un p)=
{1 \over Z} \exp\left(- {1 \over T}  H(\un q, \un p)+\frac{N}{\gamma T^2}(\tau+2\lambda T^2)J(\un q, \un p)\right).
\label{harmeasure}
\qqq
For small $\lambda$ and $\tau$,  we will see that $\rho_{\tau+2\lambda T^2}$ is a good approximation of the minimizer of the variational principle $\CF^\tau_{\lambda,N}$ at the lowest order (see also (\ref{harmeasure0})). 

\bigskip

$I^{\tau+2\lambda T^2}$ is a convex function of the parameters $(a_k,b_k,c_k)$. 
Since $\omega_0^2>0$, the contributions of each modes are independent with an hessian uniformly bounded from below in a neighborhood of the minimizer $\rho_{\tau+2\lambda T^2}$.
Any measure of the form $\mu= Z^{-1}\exp(-\frac{1}{T} \Psi) \rho_{\tau+2\lambda T^2}$, with $\Psi$ as in (\ref{psi}),  has a contribution
\begin{eqnarray}
I^{\tau+2\lambda T^2} (\mu) \ge  
C \sum_{k = - n}^{n} \big \{ a_k^2  + b_k^2  + c_k^2 \big \} 
 \label{eq: borne inf}
\end{eqnarray}
for some constant $C > 0$ independent of $N$ and $(a_k,b_k, c_k)$ close to 0.

From the exact expression (\ref{eq: K exact}), we first evaluate the difference in (\ref{itau2}) for a general Gaussian measure
(\ref{psi})
\beqnarray
K^{\tau+2\lambda T^2}(\mu) - K^{\tau}(\mu)
&=&( (\tau+2\lambda T^2)^2 - \tau^2)\frac{1}{\gamma}\sum_k\delta_k(1+2(b_k+c_k))\frac{\omega^4 }{4T^2} \sin \left( 2\pi\frac{k}{N} \right)^2 
\nonumber \\
&=& (\lambda\tau+\lambda^2 T^2)\frac{1}{\gamma }\sum_k\delta_k(1+2(b_k+c_k))\omega^4 \sin \left( 2\pi\frac{k}{N} \right)^2\,.
\label{kmutaula}
\eeqnarray
For large $N$, with the scaling $\lambda = \lambda' / N$  and $\tau = \tau' / N$,  the contribution of the $k^{th}$-Fourier mode
is at most of order ${1 \over N^2}  \sin \left( 2\pi\frac{k}{N} \right)^2$. 
From (\ref{eq: borne inf}), the corrections $(a_k,b_k, c_k)$ have to remain much smaller than ${1 \over N} \sin \left( 2\pi\frac{k}{N} \right)$, otherwise
the contribution of $I^{\tau+2\lambda T^2} $ would be too large as compared to $K^{\tau+2\lambda T^2} - K^{\tau}$.
This shows that at the lowest order $\rho_{\tau+2\lambda T^2}$ approximates correctly the minimizer of $\CF^\tau_{\lambda,N} $.
Plugging the expression (\ref{harmeasure}) of $\rho_{\tau+2\lambda T^2}$ in (\ref{itau2}) and then in (\ref{kmutaula}) gives at the  lowest order in $1/N$,
\begin{eqnarray*}
\CF^\tau_{\lambda,N} (\rho_{\tau+ 2 \lambda T^2}) =
 {(\lambda' \tau' + (\lambda ')^2 T^2) \over N^2 } \frac{\omega^4}{\gamma}\sum_k\frac{1}{\omega_k^2} \sin \left( 2\pi\frac{k}{N} \right)^2
+ o \left( {1 \over N} \right)\,.
\end{eqnarray*}
For large $N$, the last sum may be approximated by an integral 
\qq
\CF^\tau_{\lambda,N} (\rho_{\tau+2\lambda T^2})= {\kappa \over N} (\lambda' \tau'+ (\lambda ')^2 T^2) 
+ o \left( {1 \over N} \right)  \, ,
\qqq
where $\kappa$ is the conductivity of the model  (\ref{eq: kappa}).

\vskip1cm

\noindent
{\bf Remark.}
Note that a direct computation of the largest eigenvalue of the operator $L - \lambda J$ would have led also to $\CF^\tau_{\lambda,N}$.

\section{Conclusion}

We have analyzed the large deviations for  thermostated lattices of Hamiltonian oscillators by using the 
Donsker-Varadhan large deviation theory \cite{DonskerVaradhan, DonskerVaradhan2}.
After deriving the large deviation principle for a large class of models (Proposition \ref{prop: LD thme}), we analyzed the Donsker-Varadhan functional (\ref {lemma1}) and showed in section \ref{subsec: Entropy production} that it provides a variational characterization of the stationary measure which generalizes the  minimum entropy production principle. 

The current large deviation function has been computed exactly (Proposition \ref{pro: gaussian}) for thermostated lattices of harmonic oscillators on a ring.  The central issue in this computation is the evaluation of the ``dynamical activity" $K$ of the stationary measures of systems out of equilibrium.  In this paper, we used the fact that the explicit form of this measure is known in the model treated. It would be interesting to generalize these computations to Gaussian self-consistent reservoirs for which the temperatures of the heat baths vary spatially.  Except for the lack of translation invariance, this model is very similar to the one considered here.   A very challenging issue would be to investigate the current large deviations of the chain of non-harmonic oscillators.  Our analysis of the Donsker-Varadhan functional applies to this type of dynamics, but to derive quantitative estimates of the large deviations of the current, some type of approximation seems necessary.  In the approximation 
\cite{Lefevere,LefevereSchenkel,AokiSpohn,BK,Spohnphonon}, the action of the  nonlinearity of the dynamics is essentially reduced to the action of Gaussian stochastic heat baths, thus one should be able to proceed in similar fashion as in the model studied in the previous section.

\section*{Appendix}

In this appendix, we prove Proposition \ref{prop: LD thme} for the
dynamics  (\ref{dynamics mixte 0}). The generator of this dynamics is given by $L = L_S+L_H+L_{\un \gl}$, with
\begin{eqnarray*}
L_S &=& \sum_{i=1}^N  - \gamma_i p_i \frac{\partial}{\partial p_i} + \gamma_i T_i \frac{\partial^2}{\partial p_i^2} \, , 
\qquad 
L_H = \sum_{i=1}^N -\frac{\partial H}{\partial q_i}\frac{\partial}{\partial p_i}+\frac{\partial H}{\partial p_i}\frac{\partial}{\partial q_i} \, , \\
L_{\un \gl} &=& -\sum_{i=1}^N \hf T_i(\gl_{i-1}U'(q_{i-1}-q_i)+\gl_iU'(q_{i}- q_{i+1}))\frac{\partial}{\partial p_i},
\quad {\rm with} \quad \gl_0 = \gl_N =0 \, ,
\end{eqnarray*}
and open boundary conditions ($q_0=q_{N+1}=0$). The case of periodic boundary conditions can be treated similarly.

 The proof heavily relies on corollary 2.2 of the paper \cite{Wu} where it is proven that 
the large deviation principle applies provided that one can find a function $\Psi \ge 1$ such that the function
\begin{eqnarray}
\label{eq: Phi}
\Phi (\un p, \un q) = - {L \Psi \over \Psi}  (\un p, \un q)
\end{eqnarray}
diverges at infinity. This can be understood as follows. From (\ref{eq: Phi}), we deduce that $\Psi$ is an eigenvector for the operator $L + \Phi$, thus the Feynman-Kac formula implies that  for any time $t$ and initial data $(\un p(0), \un q(0))$ 
\begin{eqnarray}
&& \Psi (\un p(0), \un q(0)) = P \left(  \Psi (\un p(t), \un q(t)) \exp \left( \int_0^t \, ds  \Phi (\un p(s), \un q (s)) \right) \right) \nonumber \\
&& \qquad \qquad \qquad \qquad 
\ge  P \left( \exp \left( \int_0^t \, ds  \Phi (\un p(s), \un q (s)) \right) \right)
\label{eq: tightness}
\end{eqnarray}
where we used that $\Psi \ge 1$ in the last inequality.  This ensures the exponential integrability of $\int_0^t \, ds  \Phi (\un p(s), \un q (s))$ uniformly in time. Since $\Phi$ diverges at infinity, we recover tightness properties and the existence of an invariant measure for the dynamics (see Theorem 3.1 in \cite{Wu}).

\bigskip

\noindent
{\it Proof.}
Following \cite{Wu}, we introduce $\Psi (\un p, \un q) = \exp( F(\un p, \un q) -
 \inf F)$ with
\begin{eqnarray}
F= {1 \over 2} \hat H(\un p,\un q) + b \sum_{i=1}^N   q_i  p_i \, ,
\label{eq: test function}
\end{eqnarray}
with $b$ a small constant and $\hat H$ given by (see (\ref{ref0}))
\begin{eqnarray*}
\hat H(\un p,\un q) = \sum_{i=1}^N \frac{p_i^2}{2 T_i}+ {1 \over T_i} V(q_i)+  {
U(q_i-q_{i+1}) + U(q_{i-1} - q_i) \over 2 T_i} \, . 
\end{eqnarray*}

From (\ref{eq: Phi}), we get
\begin{eqnarray}
\label{eq: Phi1}
\Phi (\un p, \un q) = - L F  (\un p, \un q) -  \sum_{i=1}^N \gamma_i T_i  (\partial_{p_i} F)^2.
\end{eqnarray}
We compute each terms 
\begin{eqnarray*}
&& L  \hat H  (\un p, \un q) = \sum_{i=1}^N \gamma_i  - {\gamma_i \over T_i} p_i^2 
-\sum_{i=1}^{N-1} \left( {1 \over 2 T_i} -  {1 \over 2 T_{i+1}}\right)   U^\prime (q_i -q_{i+1}) (p_i+p_{i+1}) \\
&& \qquad \qquad \qquad - \sum_{i=1}^N {1 \over 2} \big( \theta_i U^\prime (q_i -q_{i+1}) + \theta_{i-1} U^\prime (q_{i-1} - q_i) \big)  p_i \, ,
\end{eqnarray*}
where $L_H \hat H$ was already computed in  (\ref{eq: LH hat H}). One has also
\begin{eqnarray*}
&& L q_i p_i = - \gamma_i p_i q_i + p_i^2   -  q_i  \partial_{q_i} H (q) 
- {T_i \over 2} \big(\theta_i  U^\prime (q_i-q_{i+1}) + \theta_{i-1}  U^\prime (q_{i-1}-q_i) \big) q_i \\
&& (\partial_{p_i} F)^2 = \left( {1 \over 2 T_i}  p_i + b q_i  \right)^2 \, . \\
\end{eqnarray*}
Combining the previous results leads to
\begin{eqnarray}
&& \Phi (\un p, \un q) =  \sum_{i=1}^N  \left(  {\gamma_i \over 4 T_i} - b \right) p_i^2
  + b q_i  \partial_{q_i} H (q)-  b^2 \gamma_i T_i q_i^2 - {\gamma_i \over 2} \nonumber \\
&& \qquad \qquad  
+ {1 \over 2} \big( \gl_i U^\prime (q_i - q_{i+1}) + \gl_{i-1} U^\prime (q_{i-1}- q_i) \big) 
\left( b T_i q_i + {p_i \over 2} \right) \nonumber \\
&& \qquad \qquad + {1 \over 4} \sum_{i=1}^{N-1} \left( {1 \over T_i} - {1 \over T_{i+1}}\right)   U^\prime (q_{i+1}-q_i) (p_i+p_{i+1}) \, ,
\label{eq: Phi 3} 
\end{eqnarray}
Since $U$ and $V$ are convex, one gets
\begin{eqnarray*}
 \sum_{i=1}^N      q_i  \partial_{q_i} H (q)
 \geq  \sum_{i=1}^N  V (q_i)+ {U(q_i-q_{i+1}) + U(q_{i+1} - q_i) \over 2} - V(0) - U(0)\, .
\end{eqnarray*}
Combining this inequality and (\ref{eq: Phi 3}), we get 
\begin{eqnarray}
\label{eq: Phi 4} 
&& \Phi (\un p, \un q) \geq   \sum_{i=1}^N  \left(  {\gamma_i \over 4 {\cal{T}_+}} - b \right) p_i^2
  + b  V (q_i) + b U(q_i-q_{i+1})  - b V(0) - b U(0) - {\gamma_i \over 2}  \nonumber \\
&& \qquad \qquad -  b^2 ( \gamma_i {\cal{T}_+}  +  {\cal{T}_+} ^2) q_i^2 
- \left( {\Delta  \over {\cal{T}}_-^2} + \gl^2 \right)  \big( U^\prime (q_{i+1}-q_i)^2 + p_i^2 \big) \, , 
\end{eqnarray}
where $\Delta = \max_i | T_i - T_{i +1}|$, ${\cal{T}_+} = \max_i  T_i $, ${\cal{T}_-} = \max_i  T_i $   and  $\gl = \max_i  |\gl_i|$.

From the assumptions (\ref{eq: assump pot}) on the potentials and (\ref{eq: Phi 4}), we see that 
as soon 
\begin{eqnarray}
\inf_i \left\{ {\delta \over  \gamma_i {\cal{T}_+}  +  {\cal{T}_+} ^2}  , {\gamma_i  \over 8 {\cal{T}_+}  } \right\} >  b
\quad {\rm and} \quad 
\inf_i  \left \{ b \delta  , {\gamma_i  \over 8 {\cal{T}_+}  } \right\} >  
{\Delta \over {\cal{T}}_-^2} +  \gl^2  ,
\end{eqnarray}
then
\begin{eqnarray}
\Phi (\un p, \un q) \geq \sum_{i=1}^N \;  c_1 \big[  p_i^2  +  V (q_i)+ U(q_i-q_{i+1})  \big]  - c_2 \, , 
\label{eq: lower phi}
\end{eqnarray}
for some constants $c_1>0$, $c_2$.
This is enough to conclude that $\Phi$ diverges at infinity and that the large deviation principle follows from (\ref{eq: tightness}) (see \cite{Wu}). 

\bigskip

From (\ref{eq: lower phi}), one can also derive an upper bound on the total current (\ref{eq: current})
\begin{eqnarray}
\Phi (\un p, \un q) + c_2 \geq \alpha N \Big| J (\un q,\un p)  \Big|  \, ,
\end{eqnarray}
for some constant $\alpha >0$. From (\ref{eq: tightness}), we deduce the exponential bound (\ref{eq: tightness current}) on the current.

\vskip1cm

\noindent
{\it Acknowledgments:}  We thank Christian Maes for useful discussions. The authors acknowledge the support of the ANR {\it LHMSHE}.

\end{document}